# Orthogonality of Diffractive Deep Neural Networks

**Shuiqin Zheng[1,2], Xuanke Zeng[1], Lang Zha[1], Huancheng Shangguan[1], Shixiang Xu[1,*], and Dianyuan Fan[2]**

[1]*Shenzhen Key Laboratory of Micro-Nano Photonic Information Technology, College of Electronic Science and Technology, Shenzhen University, Shenzhen 518060, China*

[2]*SZU-NUS Collaborative Innovation Center for Optoelectronic Science & Technology, Key Laboratory of Optoelectronic Devices and Systems of Ministry of Education and Guangdong Province, Shenzhen University, Shenzhen 518060, China*

Several laws are found for the Diffractive Deep Neural Networks (D²NN). They reveal the inner product of any two light fields in D²NN is invariant and the D²NN act as a unitary transformation for optical fields. If the output intensities of the two inputs are separated spatially, the input fields must be orthogonal to each other. These laws imply that the D²NN is not only suitable for the classification of general objects but also more suitable for applications aim to the optical orthogonal modes. Additionally, our simulation shows D²NN do well in applications like mode conversion, mode multiplexer, and optical mode recognition.

As the fastest-growing machine learning methods, deep learning [1, 2] uses multi-layered artificial neural networks and has made major advances in many domains [3-8]. Due to the parallel ability and high speed of light, beginning with the first optical realization of artificial neural networks [9, 10], the attempts of optical neural networks never stop [11-21]. Recently, a framework called diffractive deep neural network (D²NN) [22] shows the ability for the realization of the depth learning. The system using multiple optimized diffractive elements to diffract coherent fields carried patterns, and focus them into spatial regions by the classes, realizing classification effectively.

This paper finds there are several irresistible laws in D²NN. According to them, the inner product of any two light fields at any position in D²NN is invariant, which make the D²NN works as a unitary transformation to optical fields. If the output intensities are spatially separated entirely, the input optical fields must be orthogonal to each other. Therefore, we believe that the D²NN is not only applicable to the classification of general objects but more applicable for the optical orthogonal modes. Our simulations show D²NN performs well in mode conversion, mode multiplexer/ De-multiplexer and optical mode recognition.

D²NN bases on phase modulations originated by the multiple layer diffractive surfaces and diffraction propagations between the layers. So, in a D²NN system, there exist only two operations: phase modulation and propagation. In this paper, the scalar coherent light fields are regarded as vectors in Hilbert space. According to the angular spectrum theory, the propagation of an optical field in free space can be described as

$$E_z(k_x,k_y) = E(k_x,k_y)\exp(iz\sqrt{k^2-k_x^2-k_y^2}), \quad (1)$$

where $E(k_x, k_y)$ is the angular spectrum of the optical field before propagation, and $E_z(k_x, k_y)$ for the angular spectrum after propagation of a distance $z$. The propagation operator $\hat{P}$ can be defined as

$$\hat{P}E = \hat{F}^{-1}\exp(iz\sqrt{k^2-k_x^2-k_y^2})\hat{F}E(x,y), \quad (2)$$

where $\hat{F}$ is the operator for 2D-Fourier transform, correspondingly, $\hat{F}^{-1}$ represents the inverse 2D-Fourier transform. $k$ is the angular wavenumber of the fields. For any two arbitrary optical fields $Q$ and $W$ before propagation, the fields spread to any distance $z$ and evolve into $Q_z$ and $W_z$, then the inner product of $Q_z$ and $W_z$ can be deduced as

$$\begin{aligned}
\langle \hat{P}W, \hat{P}Q \rangle &= \langle W_z, Q_z \rangle \\
&= \iint W_z^*(x,y) \cdot Q_z(x,y)dxdy \\
&= \iint W_z^*(k_x,k_y) \cdot Q_z(k_x,k_y)dk_xdk_y \\
&= \iint \{W(k_x,k_y)\exp(iz\sqrt{k^2-k_x^2-k_y^2})\}^* \cdot \\
&\quad \{Q(k_x,k_y)\exp(iz\sqrt{k^2-k_x^2-k_y^2})\}dk_xdk_y \\
&= \iint W^*(k_x,k_y)\exp(-iz\sqrt{k^2-k_x^2-k_y^2}) \cdot \\
&\quad Q(k_x,k_y)\exp(iz\sqrt{k^2-k_x^2-k_y^2})dk_xdk_y \\
&= \iint W^*(k_x,k_y) \cdot Q(k_x,k_y)dk_xdk_y \\
&= \iint W^*(x,y) \cdot Q(x,y)dxdy \\
&= \langle W, Q \rangle. \quad (3)
\end{aligned}$$

Eq. (3) uses the generalized Parseval theorem and indicates that the inner product of any two light wave fields in free space is invariant during propagation. Operator $\hat{M}$ for phase modulation is defined as

$$\hat{M}E = \exp(i\phi(x,y))E(x,y). \quad (4)$$

where $\phi(x,y)$ is the phase induced by the diffractive surface. And it is easy to deduce that

$$\langle \hat{M}W, \hat{M}Q \rangle = \langle W, Q \rangle. \quad (5)$$

Eq. (5) shows the inner product of the two fields is also not changed after the phase modulation. Since there are only propagations and phase modulations in D²NN, the inner product of any two optical fields at any position in D²NN is constant, it is no exception on the detection plane. In other words, the inner product on the input plane is the same as on the detection plane. Therefore, the operation of D²NN to an optical field is a unitary transformation. We'd like to define two similarity coefficients. The first one is the field similarity coefficient. The projection from the unit vector along $Q$ to the unit vector along $W$ is

$$p_{QW} = \langle e_W, e_Q \rangle = \frac{\langle W, Q \rangle}{\sqrt{\langle Q,Q \rangle \langle W,W \rangle}}. \quad (6)$$

where $e_Q$ and $e_W$ are the unit vectors along $Q$ and $W$. And the field similarity coefficient $F_{QW}$ is defined as the square of the norm of $p_{QW}$ as

$$F_{QW} = |p_{QW}|^2 = p_{QW}^* p_{QW} = \frac{\langle Q,W \rangle \langle W,Q \rangle}{\langle Q,Q \rangle \langle W,W \rangle}. \quad (7)$$

And $F_{QW}$ is a real value between 0 and 1, representing the similarity between the fields $Q$ and $W$. If $F_{QW} = 0$, $Q$ and

$W$ are orthogonal. According to the invariant inner product in D²NN, the field similarity coefficient is also constant no matter how we optimize the D²NN. However, the detectors sense the amplitude of light, and the similarity of the fields do not mean the similarity of amplitudes. So we need the other coefficient to describe the similarity of amplitudes. Operator $\hat{A}$ denotes amplitude operator as

$$\hat{A}E = |E(x,y)| \qquad (8)$$

Then the amplitude similarity coefficient between $Q$ and $W$ can be defined as

$$I_{QW} = \frac{\langle \hat{A}W, \hat{A}Q\rangle \langle \hat{A}Q, \hat{A}W\rangle}{\langle \hat{A}Q, \hat{A}Q\rangle \langle \hat{A}W, \hat{A}W\rangle} = \frac{\langle \hat{A}W, \hat{A}Q\rangle \langle \hat{A}Q, \hat{A}W\rangle}{\langle Q, Q\rangle \langle W, W\rangle} \qquad (9)$$

And $I_{QW}$ is also a value between 0 and 1, but representing the similarity of amplitude distributions between the $Q$ and $W$. If $I_{QW} = 0$, then intensity distributions of $Q$ and $W$ are entirely separated. And it can deduce that

$$1 \geq I_{QW} \geq F_{QW} \geq 0 \qquad (10)$$

From proof above, we can get several irresistible laws.

**Law. 1**. The inner product of any two optical fields is invariant after the process of D²NN.

**Law. 2**. If $I_{QW} = 0$, then $F_{QW} = 0$, which means if the outputs are well separated, the fields must be orthogonal.

**Law. 3**. Amplitude similarity coefficient on detection plane ranges from field similarity coefficient to 1.

From Law. 3, it seems that the D²NN is unsuitable to classify patterns with high similarity. Such as the characters '6' and 'E' shown in Fig.1, the field similarity coefficient $F_{6E} \approx 0.90$. So no matter how we optimize the D²NN, we cannot make the amplitude similarity coefficient $I_{6E}$ less than 0.90, which mean the tiny intensity distribution difference on detection plane will affect the judgment of classification.

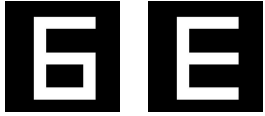

FIG. 1.The characters '6' and 'E'

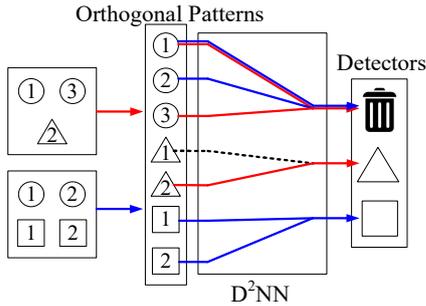

FIG. 2. How D²NN distinguish non-orthogonal patterns

However, in fact, the D²NN can do a good job for classification [22]. We think the D²NN classifies patterns based on the following mechanism and obeying the Laws. Assuming each pattern can be expressed as the sum of common patterns between different types and the sum of feature patterns peculiar to different types. Pattern $X(i,j)$ can be expressed approximately as

$$X(i,j) \approx \sum_m a_m C(m) + \sum_n b_n D_i(n). \qquad (11)$$

Here $i$ and $j$ are the type and the order of the pattern, $C$ is a set for common patterns, while $D_i$ is a set for feature patterns exclusive to type $i$. And each pattern in $\{C, D_i\}$ is normalized and orthogonal to each other. As shown in Fig.2, the circles are for the common patterns and triangles and squares represent the different type feature patterns. Obviously, the components of common patterns are useless for classification, but the components cannot be reduced for no reason and need places to settle them. So, the D²NN is optimized for finding the common patterns and feature patterns, and focus the feature patterns to their corresponding detectors, while the common patterns can be transferred to the background or the space without detectors.

From the above discussion, it can see D²NN is more suitable for orthogonal patterns. Although there is often a lack of orthogonal patterns in life, there is often having orthogonal patterns in lasers and fiber optics. For example, Laguerre Gaussian (LG) modes and Hermite Gaussian (HG) modes with different orders are orthogonal. Therefore, it is possible to use D²NN for the applications aimed to lasers modes or fiber modes, such as mode conversion, mode multiplexer/De-multiplexer, and optical mode recognition. When the inputs are orthogonal fields, Law. 1 shows the outputs are also orthogonal. So, D²NN can work as a mode converter between two sets of orthogonal modes. For this situation, the modes should be one-to-one connections. And the system is reversible and the connection rules are arbitrary, as shown in Fig.3. In order to verify these, a Matlab-coded optimizing tool is developed and available online [23].

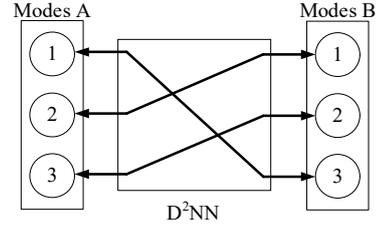

FIG. 3. Mode conversion with D²NN

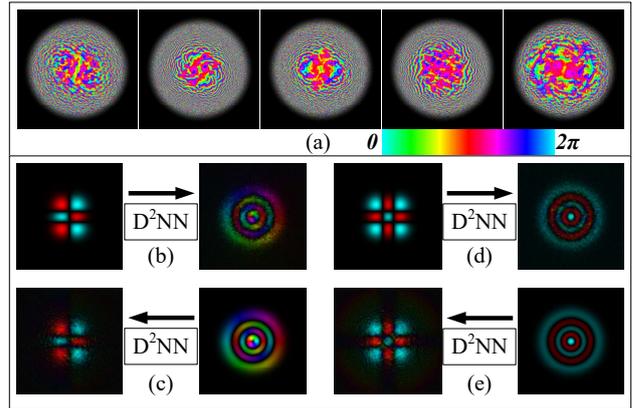

FIG. 4. A 5 layers D²NN for HG-LG mode conversion: Phase modulation of each layer (a); Interconversion between HG$_{1,2}$ and LG$_{1,3}$ (b-c); Interconversion between HG$_{2,2}$ and LG$_{0,4}$ (d-e);

Fig. 4 shows a 5 layers D²NN to transfer HG$_{m,n}$ modes to LG$_{n-m,n+m}$ modes. The phase modulation of each layer is shown in Fig. 4(a), and the distances of adjacent layers are both 30 cm. The $m$ and $n$ of HG modes are both range from 0 to 2. As shown in Fig.4 (b) and 4(c), the HG$_{1,2}$ mode converts to the LG$_{1,3}$ mode with forward-propagating, while LG$_{1,3}$ converts to HG$_{1,2}$ with backward-propagating.

And Fig. 4(d) and 4(f) show the interconversion between $HG_{2,2}$ and $LG_{0,4}$. And the conversion efficiency reached 73%, and it can up to 95% with a 10 layer design.

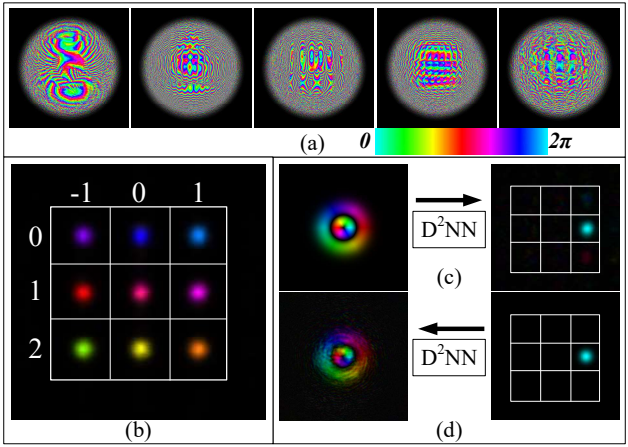

FIG. 5. A 5 layers $D^2NN$ for LG mode multiplexer: Phase modulation of each layer (a); Each LG mode is mapped to a Gaussian mode of a different position, horizontal sorting by topological orders and vertical sorting by radial orders (b); Interconversion between $LG_{11}$ mode and the Gaussian mode on the corresponding position (c-d);

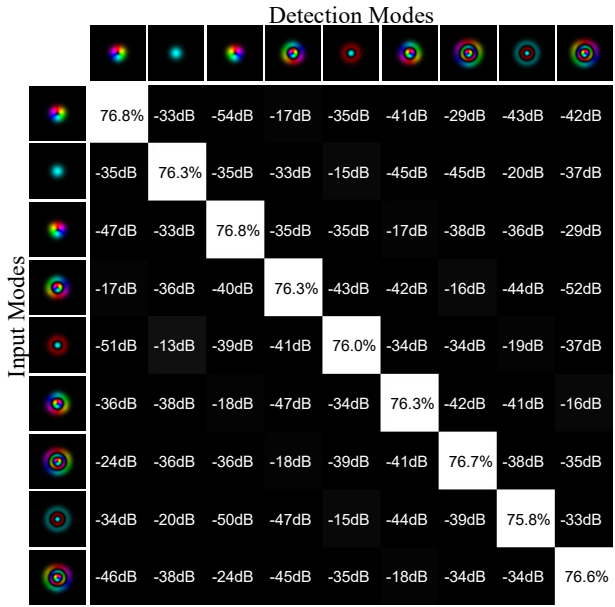

FIG. 6. Efficiency and crosstalk of the $D^2NN$ mode multiplexer

Law. 2 reveals the well spatially separated fields are also orthogonal fields. So, it is possible for transforming space overlapping orthogonal modes into spatially separated modes. It means we can optimize the $D^2NN$ for modes multiplexer or De-multiplexer. Such as the inputs are LG modes, and expected outputs are Gaussian modes at different positions. And the Gaussian modes are well separated to make them orthogonal. Fig. 5 shows a 5 layers $D^2NN$ for converting 9 LG modes to Gaussian modes on different positions. The phase modulation of each layer is shown in Fig. 5(a), and the distances between layers are also 30cm. The radial orders and topological orders of LG modes are range from 0 to 2 and -1 to 1. Each mode is mapped to a Gaussian mode of a different position, as shown in Fig. 5(b). Fig.5(c) and 5(d) show the interconversion between $LG_{11}$ mode and the Gaussian mode on the corresponding position. As shown in Fig .6, the conversion efficiency of each mode is about 76%, and the maximum crosstalk is -13dB. By using a 10 layers design, the conversion efficiency can up to 94% and the maximum crosstalk can go to -32dB.

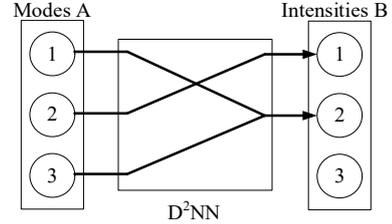

FIG. 7. Modes can be linked to a same intensity distribution

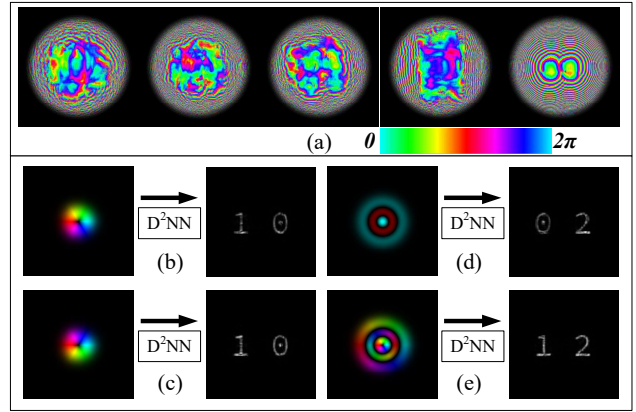

FIG.8. A 5 layers $D^2NN$ for LG mode order recognizer: Phase modulation of each layer (a); Orders show on the detection plane for $LG_{-1,0}$, $LG_{1,0}$, $LG_{0,2}$, and $LG_{1,2}$ (b-e);

According to Law. 3, when the input is a collection of orthogonal modes, amplitude similarity coefficient ranges from 0 to 1. Which means the output intensity of each input can be arbitrarily defined. Even the multi-modes can be linked to a same intensity distribution. But the connections are not reversible any more, as shown in Fig.7. Fig.8 shows a 5 layers LG mode order recognizer for showing the value of $|l|$ and $p$ of $LG_{l,p}$ modes on the detection plane. Basing on this rule, $LG_{-1,0}$ and $LG_{1,0}$ convert to the same intensity distribution in the detection plane, as shown in Fig.8(b) and 8(c). The recognizer also shows the correct value for $LG_{0,2}$ and $LG_{1,2}$, as shown in Fig.8(d) and 8(e).

In summary, we find there are several laws in $D^2NN$. First, the inner product of any two light fields in $D^2NN$ is invariant and the $D^2NN$ act as a unitary transformation for optical fields. Then, the laws reveal that if the output intensities of the two inputs are entirely separated, the input fields must be orthogonal. Last, the laws show amplitude similarity coefficient in the detection plane ranges from field similarity coefficient to 1. Basing on these laws, we believe that the $D^2NN$ is not only suitable for the classification of general objects but more applicable for applications of optical orthogonal modes. And our simulation shows $D^2NN$ do well in applications like mode conversion, mode multiplexer/De-multiplexer, and optical mode recognition. We believe $D^2NN$ is a powerful tool for manipulation of optical orthogonal modes, which have the potential applications in many fields, especially in modal characterization and optical communication.

We acknowledge the supports of National Natural Science Funds of China (61490710, 61775142 and 61705132), The Science and Technology Planning Project of Guangdong Province (2016B050501005), The Specialized Research Fund for the Shenzhen Strategic Emerging Industries Development (JCYJ20170412105812811), and the fund of The International Collaborative Laboratory of 2D Materials for Optoelectronics Science and Technology, Shenzhen University (2DMOST2018019).
* shxxu@szu.edu.cn